\documentclass[aps,prd,preprint,preprintnumbers,unsortedaddress,superscriptaddress,showpacs,nofootinbib]{revtex4-1}

\pdfoutput=1

\usepackage{graphicx}
\usepackage{amsmath}
\usepackage{amsfonts}
\usepackage{amssymb}
\usepackage{color}

\usepackage[colorlinks, citecolor=blue,anchorcolor=red,menucolor=red, linkcolor=red,filecolor=red,runcolor=red,urlcolor=blue,frenchlinks=red]{hyperref}

\begin{document}
\arraycolsep1.5pt


\title{ Anomalous enhancement of the isospin-violating $\Lambda(1405)$ production by a triangle singularity in $\Lambda_c\rightarrow\pi^+\pi^0\pi^0\Sigma^0$}


\author{L.~R.~Dai}
\email{dailr@lnnu.edu.cn}
\affiliation{Department of Physics, Liaoning Normal University, Dalian 116029, China}
\affiliation{Departamento de F\'isica Te\'orica and IFIC, Centro Mixto Universidad de Valencia-CSIC,
Institutos de Investigac\'ion de Paterna, Aptdo. 22085, 46071 Valencia, Spain
}

\author{R. Pavao}
\email{rpavao@ific.uv.es}
\affiliation{Departamento de F\'isica Te\'orica and IFIC, Centro Mixto Universidad de Valencia-CSIC,
Institutos de Investigac\'ion de Paterna, Aptdo. 22085, 46071 Valencia, Spain
}

\author{S. Sakai}
\email{shuntaro.sakai@ific.uv.es}
\affiliation{Departamento de F\'isica Te\'orica and IFIC, Centro Mixto Universidad de Valencia-CSIC,
Institutos de Investigac\'ion de Paterna, Aptdo. 22085, 46071 Valencia, Spain
}

\author{E.~Oset}
\email{Eulogio.Oset@ific.uv.es}
\affiliation{Departamento de F\'isica Te\'orica and IFIC, Centro Mixto Universidad de Valencia-CSIC,
Institutos de Investigac\'ion de Paterna, Aptdo. 22085, 46071 Valencia, Spain
}

\date{\today}
\begin{abstract}
 The decay of $\Lambda_c^+$ into $\pi^+\pi^0\Lambda(1405)$ with the
 $\Lambda(1405)$ decay into $\pi^0\Sigma^0$ through a triangle diagram
 is studied.
 This process is initiated by  $ \Lambda_c^+ \to \pi^+ \bar{K}^*N $, then the $\bar{K}^*$ decays into $\bar{K} \pi$ and
 $\bar{K} N$  produce the $\Lambda(1405)$ through a triangle loop containing $\bar{K}^*N\bar{K}$ which develops
 a singularity around $1890$~MeV. This process is prohibited by the isospin symmetry, but  the decay into this channel is enhanced by the contribution of the
 triangle diagram, which is sensitive to the mass of the internal
 particles.  We find a narrow peak in the $\pi^0\Sigma^0$ invariant mass
 distribution, which originates from the $\Lambda(1405)$ amplitude, but  is tied to the  mass differences between the charged and neutral $\bar{K}$ or $N$ states.
  The observation of the unavoidable peak of the triangle singularity in
 the isospin-violating $\Lambda(1405)$ production would provide further
 support for hadronic molecular picture of the $\Lambda(1405)$  and further information on the $\bar{K} N $ interaction.
\end{abstract}

\maketitle

\section{Introduction}
\label{sec_introduction}

The possible role of a triangle singularity (TS) in hadronic reactions has been
studied for a long time. The TS appears from a loop contribution in the decay of a particle
$1$ into two particles $2$ and $3$ through the following process:
at first the particle $1$ decays into particles $A$ and $B$, and the
particle $A$ subsequently decays into particles $2$ and $C$, and finally
the particles $B$ and $C$ merge and form the particle $3$ in the final state.
The TS was originally studied in Ref.~\cite{Landau:1959fi}, and
it was found in Ref.~\cite{Coleman:1965xm} that
the TS appears when the loop process has a classical counterpart, $i.e.$,
all the momenta of the particles in the loop (the particles $A$, $B$ and $C$ in the above
reaction) can be placed  on-shell and the momenta of the
particles $2$ and $B$ are antiparallel.
A refined formulation based on Feynman diagrams and a simple formula
for the position of the TS were given in
Ref.~\cite{Bayar:2016ftu}.
One should note that the singularity would be smeared by the width of
hadrons and appear as a broad peak in actual reactions.
This peak purely comes from a kinematical effect, then we cannot
associate this peak with a resonant state.
In Refs.~\cite{Liu:2015taa,Ketzer:2015tqa,Aceti:2016yeb}, it was found that the
``$a_1(1420)$'' peak in the $\pi f_0(980)$ invariant mass distribution
with a $p$-wave pion observed by COMPASS Collaboration
\cite{Adolph:2015pws} can be understood as a peak of a TS.
In this process, the triangle diagram is formed by
the $a_1(1260)$ decaying  into $K\bar{K}^*$ ($K^*\bar{K}$)
 with a subsequent $\bar{K}^* \to \pi\bar{K}$ ($K^* \to \pi K$ ) decay and merging $K\bar{K}$ to form
$f_0(980)$.
The $a_1(1260)$ and the $f_0(980)$ have sizable couplings to the
$K\bar{K}^*+ c.c.$ and the $K\bar{K}$ channels, respectively, because they are dynamically
generated through the coupled-channel effect of hadrons as studied
in Refs.~\cite{Roca:2005nm,lutz,geng} and
Refs.~\cite{Oller:1997ti,Locher:1997gr,Kaiser:1998fi,Nieves:1999bx,Gamermann:2006nm,Liang:2014tia,Xie:2014tma}
for the $a_1(1260)$ and the $f_0(980)$, respectively.
Actually, the large coupling of the internal particles and the final-state hadron
is crucial to have a prominent peak of the TS.
Nowadays, many hadronic molecular states have been studied as summarized
in Ref.~\cite{Guo:2017jvc}.
Because these states tend to have a large coupling to their
constituent hadrons, the observation of the inevitable peak from the TS
would provide further clues to clarify the nature of the hadronic
molecules.
Other than the ``$a_1(1420)$'', the interpretation of the
``$f_1(1420)$'' and ``$f_2(1810)$'' in the PDG \cite{pdg} as a peak of
the triangle singularity was proposed in Refs.~\cite{Debastiani:2016xgg} and
\cite{Xie:2016lvs}, respectively.
Furthermore, possible manifestations of the TS in heavy sector were investigated in
{Refs.~\cite{Szczepaniak:2015eza,Liu:2015taa,Bondar:2016pox,Pilloni:2016obd,Liu:2017vsf,Pavao:2017kcr,Sakai:2017hpg}}.

On the other hand, it was found that the TS gives significant
contribution to the isospin-violating process.
In Refs.~\cite{Wu:2011yx,Aceti:2012dj,Wu:2012pg},
the role of the triangle diagram in the unusually large isospin-violating $\pi^0f_0(980)$ production
from $\eta(1405)$ observed in BESIII \cite{BESIII:2012aa} was studied.
The triangular diagrams
formed by $K^{*-}K^+K^-$ and $\bar{K}^{*0}K^0\bar{K}^0$ contribute
to this process because of the sensitivity of the triangle singularity to the masses of
the particles in the loop diagram,
the TS can have a sizable contribution in the isospin-violating process.
It is noteworthy that the shape of the $f_0(980)$ resonance appears
narrower than observed in other processes
because the resonance shape is modified by the amplitude of the triangle
diagram, which gives the width with the order of the charged- and
neutral-kaon mass difference. Also, the line shape of the $\pi\pi$ invariant mass distribution
calculated with the triangle diagram agrees with what was observed
experimentally \cite{BESIII:2012aa}.
Following these studies, the isospin-violating $f_0(980)$ productions
enhanced by the TS in the $D_s^+\rightarrow\pi^+\pi^0f_0(980)$ and
$\bar{B}^0_s\rightarrow J/\psi\pi^0f_0(980)$ processes were studied in
Refs.~\cite{Sakai:2017iqs} and \cite{Liang:2017ijf}, respectively.

In this paper, we focus on the isospin-violating
$\Lambda_c^+\rightarrow\pi^+\pi^0\Lambda(1405)$ process with the
$\Lambda(1405)$ decay into $\pi^0\Sigma^0$ from the triangle diagram.
The triangle diagram is formed by the decay of
$\Lambda_c^+$ into $\pi^+{K}^{*-}p$ $(\pi^+\bar{K}^{*0}n)$ followed by the decay of
${K}^{*-}\rightarrow\pi^0K^-$ $(\bar{K}^{*0}\rightarrow\pi^0\bar{K}^0)$ and the fusion of
the $K^-p$ $(\bar{K}^0n)$ to form $\Lambda(1405)$.
From the formula of Eq.~(18) in Ref.~\cite{Bayar:2016ftu}, a singularity
from the triangle diagram would appear around $1890$~MeV in the
$\pi^0\Lambda(1405)$ invariant mass distribution.
The $\Lambda(1405)$ is successfully described as a hadronic molecule
\cite{Dalitz:1959dn,Dalitz:1960du,Dalitz:1967fp,Kaiser:1995eg,Oset:1997it,Oller:2000fj,Lutz:2001yb,GarciaRecio:2002td,Jido:2003cb},
and has a large coupling to the $\bar{K}N$ and
the $\pi\Sigma$ channels (see also Refs.~\cite{Hyodo:2011ur,Kamiya:2016jqc} and
references therein for the details).
The decay of heavy hadrons containing a charm or bottom
quark is an exciting field in hadron physics as summarized in
Ref.~\cite{Oset:2016lyh}, and
particularly the $\Lambda(1405)$ production
in the $\Lambda_c^+$, $\chi_{c0}(1P)$ and $\Xi_b$ decays was studied in
Refs.~\cite{Hyodo:2011js,Miyahara:2015cja}, {Ref.~\cite{liu:2017efp}} and Ref.~\cite{Miyahara:2018lud}, respectively, where the $\Lambda(1405)$
affects the $\pi\Sigma$ or $\bar{K}N$ mass distribution through the
final-state rescattering.
Considering the external $W^+$ emission for the transition of
$\Lambda_c^+$ into $\pi^+\bar{K}^*N$, which would give the main
contribution to this process, the $\Lambda(1405)$ production is isospin
forbidden. Indeed the $W$ produces the $\pi^+$ in one vertex and in the other one includes a $cs$ transition.
We have thus $\pi^+$  and $sud$, with $ud$ in $I=0$, because there these quarks are spectators. Thus the $sud$
final state has $I=0$ and hadronizes in $\bar{K}^*N$ (see Fig.~\ref{Lambdac}).
Meanwhile, the possible effect of the TS on the
$\Lambda(1405)$ production was studied in
Refs.~\cite{Wang:2016dtb,Xie:2017gwc,Bayar:2017svj}.
Now, as found in
Refs.~\cite{Wu:2011yx,Aceti:2012dj,Wu:2012pg,Sakai:2017iqs,Liang:2017ijf}
for the $f_0(980)$ production,
we expect that the isospin-violating $\Lambda(1405)$ production
is enhanced by the TS around $1890$~MeV in
the $\pi^0\Lambda(1405)$ mass distribution,
where the triangle singularity would appear from the formula in
Ref.~\cite{Bayar:2016ftu}, and that a narrow peak around the
$\Lambda(1405)$ energy in the $\pi^0\Sigma^0$ mass distribution would appear.
The observation of the TS in this isospin-violating $\Lambda(1405)$
production would give further support to the hadronic
molecular picture of the $\Lambda(1405)$ resonance, and provide us better
understanding on the triangle singularity.


\section{Formalism}
\label{sec:form}

\begin{figure}
	\begin{center}
		\includegraphics[width=0.48\textwidth]{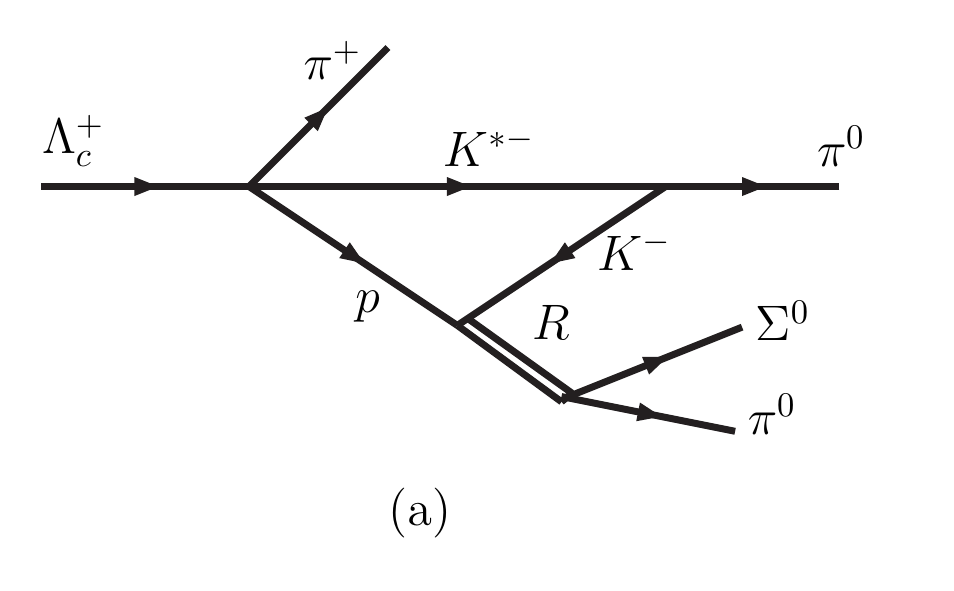}	\includegraphics[width=0.48\textwidth]{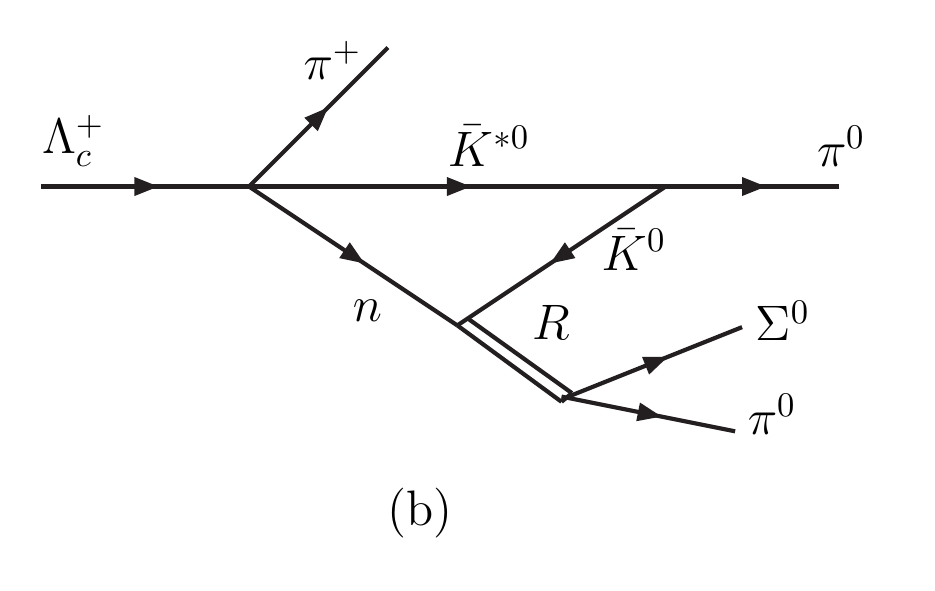}	
	\end{center}
	\caption{\label{fig:LcpiR} Diagram for the decay of $\Lambda_c^+\rightarrow \pi^+\pi^0\pi^0\Sigma^0$ }
\end{figure}

In the present study, we investigate the $\Lambda_c^+\rightarrow \pi^+\pi^0\pi^0\Sigma^0$ decays
via $\Lambda(1405)$ formation. The process of $\Lambda_c^+\rightarrow \pi^+ K^{*-} p$  followed by the
$K^{*-}$ decay into $\pi^0 K^-$  and the merging of the $K^- p$ into $\Lambda(1405)$  (see Fig. \ref{fig:LcpiR}(a))
or $\Lambda_c^+\rightarrow \pi^+ \bar{K}^{*0} n$  followed by the
$\bar{K}^{*0} $ decay into $\pi^0 \bar{K}^{0}$ and the merging of the $\bar{K}^{0} n$ into $\Lambda(1405)$
(see Fig. \ref{fig:LcpiR}(b)) generate a singularity, and we will see a signal for the   $\Lambda(1405)$  around
$1420$ MeV  because it  comes from $ \bar{K} N$ which couples to the second pole at 1420 MeV in the invariant mass of $\pi^0\Sigma^0$.
In the  study of Ref.~\cite{Oset:1997it}, the $\Lambda(1405)$  appears as the
dynamically generated state  of $K^- p$, $\bar{K}^0 n$, $\pi^0 \Lambda$, $\pi^0\Sigma^0$, $\eta\Lambda$ $\eta \Sigma^0$, $\pi^+\Sigma^-$,
$\pi^-\Sigma^+$, $K^+ \Xi^-$   and $K^0 \Xi^0$  in the coupled-channels calculation.

\begin{figure}
	\begin{center}
		\includegraphics[width=0.52\textwidth]{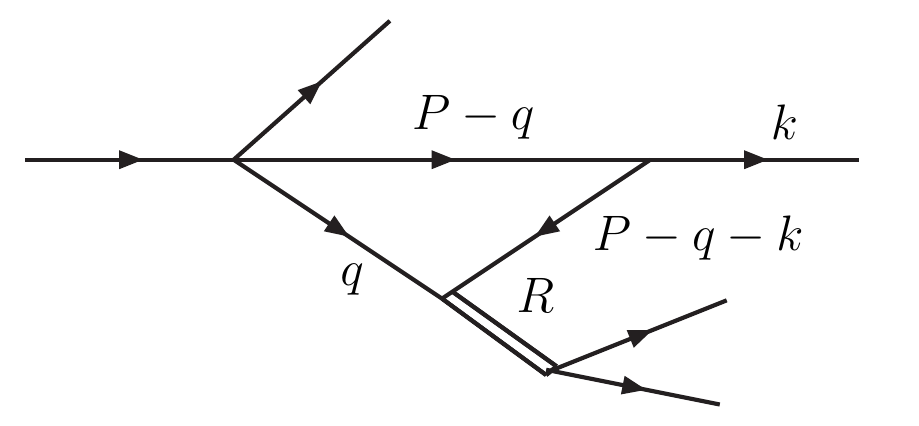}
	\end{center}
	\caption{\label{fig:mom} The momenta assignment for the decay process}
\end{figure}

We will analyze the effect of triangle singularities in the decay of  $\Lambda_c^+\rightarrow \pi^+\pi^0\pi^0\Sigma^0$.
{In this study, we focus on the decay channel of $\pi^0\Sigma^0$ from $\Lambda(1405)$, which
does not contain the $I=1$ contribution and have small $I=2$ one, to focus on the isospin violation.}
 The complete Feynman diagram for the decay with the triangle mechanism through the  $\Lambda(1405)$  baryon is shown in Fig. \ref{fig:LcpiR} and
the momenta assignment for the decay process is given in Fig.~\ref{fig:mom}.

Now we would like to evaluate the $\Lambda_c^+ \to \pi^+\pi^0 R$ with $R \to \pi^0\Sigma^0$ process which  produces the triangle diagram shown in Fig. \ref{fig:LcpiR},
where $R$ stands for the $\Lambda(1405)$ resonance.

First, let us consider the $T$ matrix of  Fig. \ref{fig:LcpiR}(a), which is given by

\begin{widetext}
\begin{equation}
\label{eq:t1}
-i t = i \sum_{\text{pol. of } K^* } \int \frac{d^4 q}{(2 \pi)^4}  \frac{i \ t_{\Lambda_c^+ \to  \pi^+ K^{*-}  p}}{q^2-M_p^2+i \epsilon} \frac{i \ t_{K^{*-}\to \pi^0 K^- } }{(P-q)^2-m_{K^{*-}}^2+i \epsilon}
\frac{i \ t_{K^- p \to \pi^0\Sigma^0}}{(P-q-k)^2 - m_{K^-}^2+i\epsilon}.
\end{equation}
\end{widetext}
The amplitude in Eq.~\eqref{eq:t1} is evaluated in the center-of-mass (CM) frame of $\pi^0 R$. Thus we need to calculate the three vertices, $t_{\Lambda_c^+ \to  \pi^+ K^{*-}  p} $, $t_{K^{*-} \to \pi^0 K^- }$
and $t_{K^- p \to \pi^0\Sigma^0}$, in Eq.~\eqref{eq:t1}.

\subsection{Decay mechanism at quark level}
\begin{figure}
\includegraphics[width=0.48\textwidth]{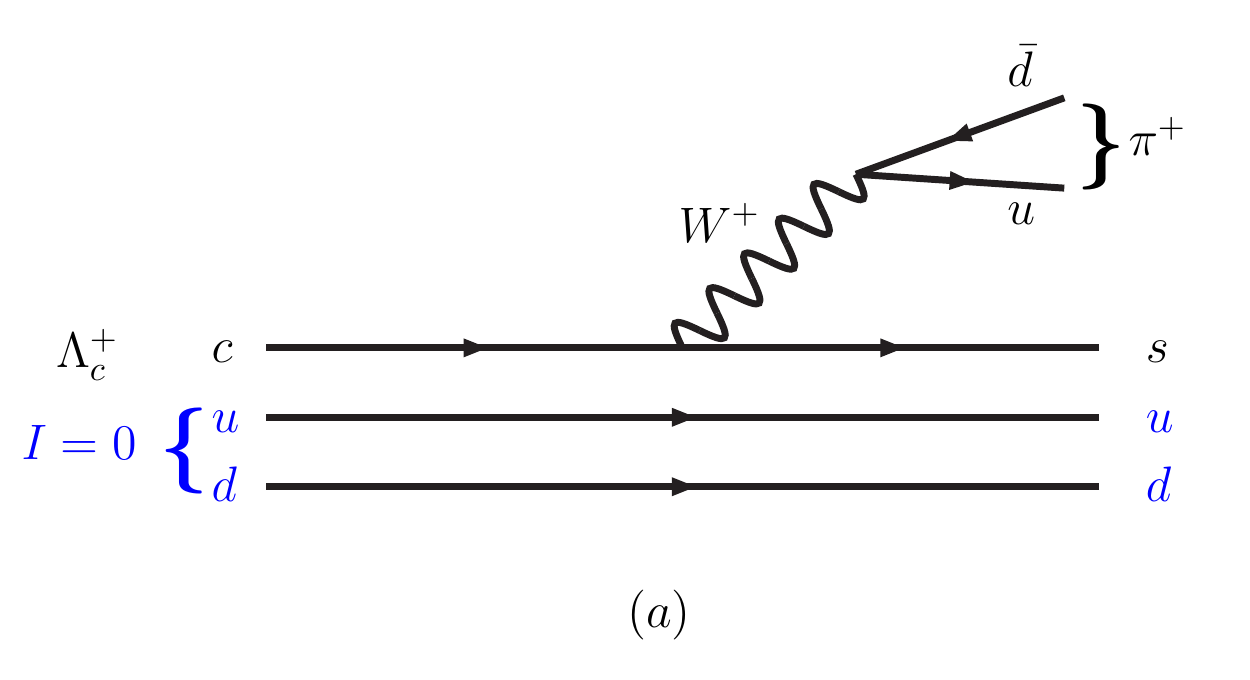} ~~\includegraphics[width=0.48\textwidth]{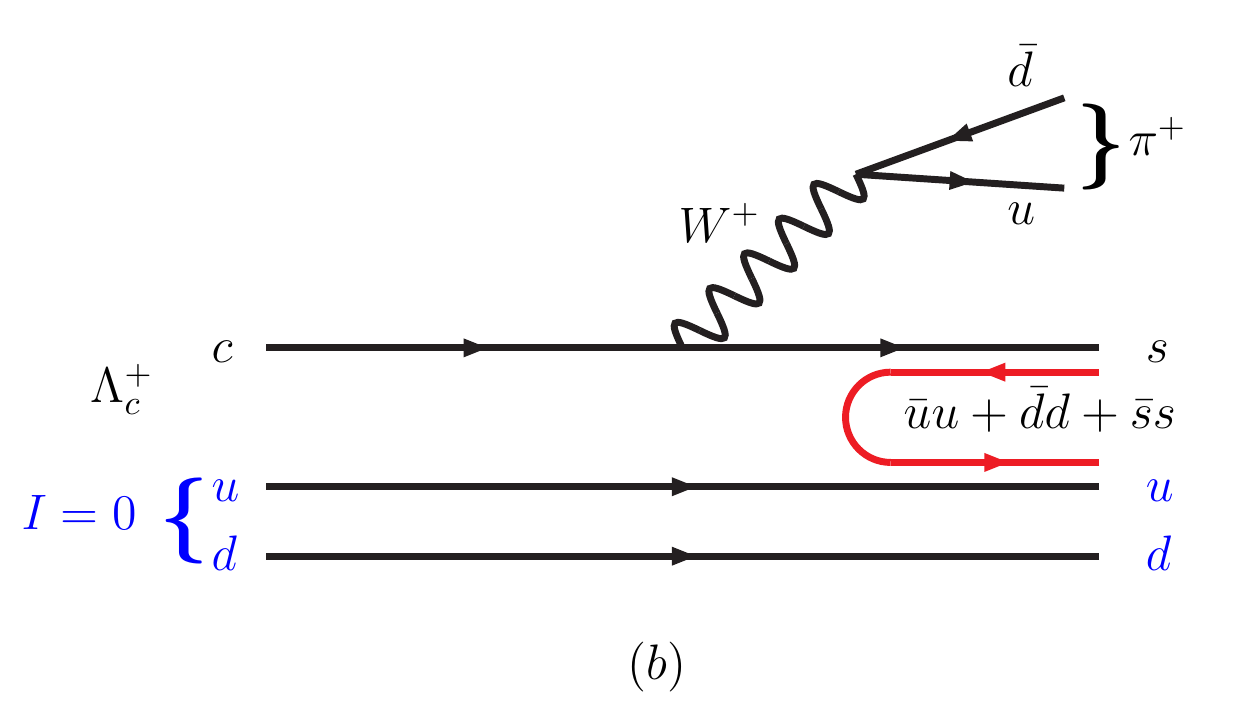}
\caption{\label{Lambdac} (a) Diagram for $\Lambda_c^+\rightarrow \pi^+\pi^0\pi^0\Sigma^0$  decay; (b) Hadronization through $\bar{q}q$
	creation with vacuum quantum numbers.}
\end{figure}
Now we look into the $\Lambda_c^+$ decay mechanism at quark level depicted in Fig.~\ref{Lambdac}(a).
At the quark level, the Cabibbo-allowed vertex is formed  through an external emission of a $W$
boson \cite{chau},  which is also color-favored, producing a $u \bar{d}$ pair that forms the $\pi^{+}$, with the remaining $ s u d$ quarks hadronizing
from a created vacuum $\bar{u}u+\bar{d}d+\bar{s}s$ state. Note that $ud$ in the $\Lambda_c^+$ are in $I=0$ and since they are spectators in the reaction they
also have $I=0$ in the final state of Fig.~\ref{Lambdac}~(a).  The resulting meson and baryon from the hadronization of Fig.~\ref{Lambdac}~(b) are easily obtained by writing
\begin{equation}\label{had}
H=\sum\limits_{i=1}^{3} s \bar{q}_i q_i \frac{1}{\sqrt{2}}(ud-du) =\sum\limits_{i=1}^{3} M_{3i} q_i \frac{1}{\sqrt{2}}(ud-du) \nonumber\, ,
\end{equation}
where $M_{ij}$ is the $q\bar{q}$ matrix with the $u,\,d,\,s$ quarks,
\begin{eqnarray}
M = \left(
           \begin{array}{ccc}
             u\bar u & u \bar d & u\bar s \\
             d\bar u & d\bar d & d\bar s\\
             s\bar u & s\bar d & s\bar s\\
           \end{array}
         \right).
\end{eqnarray}

However, we are interested in $K^*N$ production for the triangle singularity.
Thus we  write the $q\bar{q}$ matrix in terms of physical vector mesons  as
\begin{equation}
\label{eq:V}
M_{ij} \to V_{\mu} =
\left(
\begin{array}{ccc}
\frac{\rho^0}{\sqrt{2}}  + \frac{\omega }{\sqrt{2}} & \rho^+ & K^{* +} \\
\rho^- & -\frac{\rho^0}{\sqrt{2}}  + \frac{\omega}{\sqrt{2}}  & K^{* 0}  \\
K^{* -} & \bar{K}^{* 0} & \phi  \\
\end{array}
\right)_{\mu}\,
.\end{equation}
By looking at the quark content of the octet baryons in   Ref.~\cite{Miyahara} (see table III of that work),
we get
\begin{equation}\label{h}
| H \rangle = K^{*-}  p+ \bar{K}^{*0} n -\sqrt{\frac{2}{3}} \phi \Lambda\, .
\end{equation}
However, we will  neglect the $ \phi \Lambda$ component since this does not contribute to our triangle singularity mechanism.

\subsection{Calculation of the three vertices}
\subsubsection{ First vertex}
{The $\Lambda_c^+\rightarrow\pi^+ K^{*-}p$ process
can proceed via $s$-wave, and we take the amplitude of the process
$t_{\Lambda_c^+\rightarrow\pi^+K^{*-}p}$ as}
\begin{equation}\label{eq:A}
 t_{\Lambda_c^+ \to  \pi^+ K^{*-}p} = A {\vec\sigma}\cdot {\vec\epsilon}.
\end{equation}
The  $K^{*-}p$ invariant  mass distribution of the $\Lambda_c^+ \to  \pi^+ K^{*-}  p$  decay,
\begin{widetext}
\begin{equation}
\label{eq:ratio}
\frac{d \Gamma_{\Lambda_c^+ \to  \pi^+ K^{*-}p}}{d M_{\text{inv}}( K^{*-}p)}=\frac{1}{(2 \pi)^3} \frac{2M_{\Lambda_c^+} 2 M_p}{4 M_{\Lambda_c^+}^2} p_{\pi^+} \widetilde{p}_{K^{*-}}\overline{\sum} \sum \left|t_{\Lambda_c^+ \to  \pi^+ K^{*-}p} \right|^2,
\end{equation}
\end{widetext}
where $p_{\pi^+}$ is the momentum of $\pi^+$ in the $\Lambda_c^+$ rest frame, and $\widetilde{p}_{K^{*-}}$ is the momentum of $K^{*-}$ in the $ K^{*-}p$ rest frame,
\begin{subequations}
\begin{align}
&	p_{\pi^+} = \frac{\lambda^{1/2} (M_{\Lambda_c^+}^2, m^2_{\pi}, M^2_{\text{inv}}(K^{*-}p))}{2 M_{\Lambda_c}},	\\
&  \widetilde{p}_{K^{*-}}=\frac{\lambda^{1/2} (M^2_{\text{inv}}(K^{*-}p), m^2_{K^{*-}}, M^2_{p})}{2 M_{\text{inv}}(K^{*-}p)},
\end{align}
\end{subequations}
with $\lambda(x,y,z)$ the ordinary K{\"a}llen function.
Now, if we square the $T$ matrix in Eq.~\eqref{eq:A} and sum and average over the polarizations and spins, we get
\begin{align}
\overline{\sum} \sum &\left|t_{\Lambda_c^+ \to  \pi^+ K^{*-}p} \right|^2=3 \left| A \right|^2.
\end{align}
Thus we get
\begin{equation}
\frac{|A|^2}{\Gamma_{\Lambda_c^+}}=\frac{Br(\Lambda_c^+ \to  \pi^+ K^{*-}  p)}{\int^{M_{\Lambda_c^+} -m_{\pi^+} }_{M_{K^{*-}} +m_{p} }\frac{3}{(2\pi)^3}\frac{M_p}{M_{\Lambda_c^+}}p_{\pi^+} \widetilde{p}_{K^{*-}} dM_{\rm inv}({K^{*-}}p) }.
\end{equation}
By calculating the width of this decay, using the experimental  branching ratio of this decay $Br(\Lambda_c^+ \to  \pi^+ K^{*-}  p)=(1.5\pm 0.5) \times 10^{-2}$ {\cite{pdg}}, we can determine the value of the constant $|A|$.

\subsubsection{Second vertex}
Now we calculate the contribution of the vertex $K^{* -} \rightarrow \pi^0 K^-$  by using the chiral invariant lagrangian with local hidden symmetry given in Refs.~\cite{LHS1,LHS2,LHS3,LHS4}, which is
\begin{equation}
\mathcal{L}_{VPP} = -i g \left < V^{\mu} \left[P, \partial_{\mu} P\right] \right >.\label{eq:vpp}
\end{equation}
The brackets $\left<...\right>$ means the trace over the SU(3) flavour matrices, and  the coupling is given by $g=m_V/2 f_{\pi}$ in the local hidden gauge, with $m_V=780 \ \text{MeV}$ and $f_{\pi}$=93 MeV.
The SU(3) matrix  for the vector mesons is given in Eq.~\eqref{eq:V} and  for  pseudoscalar  mesons is given by
\begin{equation}
P=\begin{pmatrix}
\frac{\pi^0 }{\sqrt{2}} +\frac{\eta}{\sqrt{3}}  +\frac{\eta'}{\sqrt{6}}   & \pi^+ &  K^+ \\
\pi^- & -\frac{\pi^0 }{\sqrt{2}} +\frac{\eta}{\sqrt{3}}  +\frac{\eta'}{\sqrt{6}}  & K^0\\
K^- & \bar{K}^0 & - \frac{\eta}{\sqrt{3}}+\frac{\eta'}{\sqrt{6}}
\end{pmatrix},
\end{equation}
where the standard mixing of $\eta$ and $\eta'$ has been assumed~\cite{bramon}. Performing the matrix operations and the trace we get for $ K^{* -} \to \pi^0 K^-$,
\begin{equation}
\mathcal{L}_{K^{* -} \pi^0 K^-} =-ig \frac{1}{\sqrt{2}} {K^{* -}}^{\mu} \left(\pi^0 \partial_{\mu} K^+ - \partial_{\mu} \pi^0 K^+ \right).
\end{equation}
So, for the $t$ matrix we get,
\begin{align}
\label{eq:v21}
-i t_{K^{* -} \to \pi^0 K^- } =& i g \frac{1}{\sqrt{2}} \epsilon^{\mu}_{K^{* -}}
 (p_{K^-}-p_{\pi^0})_\mu \\
 \simeq&  i g \frac{1}{\sqrt{2}} \vec{\epsilon}_{K^{* -}} \cdot (\vec{p}_{\pi^0}-\vec{p}_{K^-}),\label{eq:v21a}
\end{align}
with $\vec{p}_{K^-}$ and $\vec{p}_{\pi^0}$ calculated in the CM frame of $\pi^0 R$.
At the energy where the triangle singularity appears, compared with the mass of $K^{*-}$,  the momentum of $K^{*-}$  is small enough,
so we omit the zeroth component of the polarization vector in Eq.~\eqref{eq:v21}.


\subsubsection{Third vertex}
The third vertex corresponds to the mechanism for the production of the $\pi^0\Sigma^0$ pair in the final state, after the rescattering of the $K^- p$
that dynamically generates the $\Lambda(1405)$ resonance  as intermediate state. We will write the vertex as
\begin{equation}
\label{eq:vertex3}
-it_3=-it_{if}\ ,~~~~\\
t_3 \equiv t_{K^- p \to \pi^0\Sigma^0}
\end{equation}
where
 $t_{if}$ is the $if$ element of the $10\times 10$ scattering matrix $t$ for the channels
 $K^- p$ (1), $\bar{K}^0 n$ (2), $\pi^0 \Lambda$ (3), $\pi^0\Sigma^0$ (4), $\eta\Lambda$ (5), $\eta \Sigma^0$ (6), $\pi^+\Sigma^-$ (7),
$\pi^-\Sigma^+$ (8), $K^+ \Xi^-$  (9),  and $K^0 \Xi^0$ (10), in the coupled-channels calculation.
We have $i=1$ for the diagrams of  Fig. \ref{fig:LcpiR}(a), while the index $f$ stands for channel $4$. The $t$ matrix is obtained using the Bethe-Salpeter equation, with the tree level potentials given in Ref.~\cite{Oset:1997it}. The loop functions for the intermediate states are regularized using the cutoff method and the peak of the $\Lambda(1405)$ is well reproduced using a cutoff of $630$ MeV. We will need this parameter for the next steps of the calculation, being necessary in order to evaluate the loop integral in the diagram of Fig. \ref{fig:LcpiR}.

\subsection {The total amplitude }
Now we obtain the final amplitude of  $\Lambda_c^+\rightarrow \pi^+\pi^0 \pi^0\Sigma^0$ for Fig. \ref{fig:LcpiR}(a),
\begin{equation}
t_{\Lambda_c^+ \rightarrow \pi^+\pi^0 \pi^0\Sigma^0} = -A ~\frac{1}{\sqrt{2}} ~g~ {\vec \sigma} \cdot {\vec k} \ t_{K^- p \to \pi^0\Sigma^0}  \ t_T,
\end{equation}
where for simplicity we use $ t_T \equiv t_T(m_{K^{*-}},M_p,m_{K^-})$ for Fig. \ref{fig:LcpiR}~(a) decay, now
\begin{widetext}
\begin{equation}
t_T=i \int \frac{d^4 q}{(2 \pi)^4}   \frac{2 M_p}{q^2-M_p^2+i \epsilon} \frac{1}{(P-q)^2-m^2_{K^{*-}}+i \epsilon} \frac{1}{(P-q-k)^2 - m_{K^-}^2+i \epsilon} \ \left(2+ \frac{\vec{q} \cdot \vec{k}}{\vec{k}^2}\right).
\label{eq:tt}
\end{equation}
\end{widetext}
where the term     $\vec{q} \cdot \vec{k}/{\vec{k}^2}$  comes from the term proportional to $\vec{q}$ in the integrand, which given the fact that $\vec{k}$ is the only vector non integrated in  Eq.~(\ref{eq:tt}),
can be written as $\vec{k} (\vec{q} \cdot \vec{k}/{\vec{k}^2})$ \cite{acetiepja}.
The analytical integration of $t_T$  over $q^0$ leads to
\begin{widetext}
\begin{align}\label{eq:tt2}
 t_T =& \int \frac{d^3 q}{(2 \pi)^3} \frac{2 M_p}{8 \omega_{K^{*-}} \omega_p \omega_{K^-}} \frac{1}{k^0-\omega_{K^-}-\omega_{K^{*-}}+i\frac{\Gamma_{K^{*-}}}{2}} \frac{1}{P^0+\omega_p+\omega_{K^-}-k^0} \left(2+ \frac{\vec{q} \cdot \vec{k}}{\vec{k}^2}\right)\\\nonumber
&\times \frac{1}{P^0 - \omega_p -\omega_{K^-}-k^0+ i \epsilon}  \frac{2P^0 \omega_p + 2 k^0 \omega_{K^-} -2(\omega_p+\omega_{K^-})(\omega_p+\omega_{K^-}+\omega_{K^{*-}})}{P^0-\omega_{K^{*-}}-\omega_p+i\frac{\Gamma_{K^{*-}}}{2}},
\end{align}
\end{widetext}
with  $P^0=M_{\rm inv}(\pi^0 \Lambda(1405))$,  $\omega_p=\sqrt{\vec{q}~^2+M_p^2}$, $\omega_{K^-}=\sqrt{(\vec{q}+\vec{k})^2+m_{K^-}^2}$,
$\omega_{K^{*-}}=\sqrt{\vec{q}~^2+m_{K^{*-}}^2}$.  The energy $k^0$  and momentum $|\vec{k}|$ of $\pi^0$ emitted from $\bar{K}^*$  are given by
\begin{equation}
k^0=\frac{M^2_{\rm inv}(\pi^0 \Lambda(1405))+m_{\pi^0}^2-M^2_{\rm inv}(\pi^0 \Sigma^0)}{2 M_{\rm inv}(\pi^0 \Lambda(1405))},
\end{equation}
\begin{equation}
|\vec{k}|=\frac{\lambda^{1/2}(M^2_{\rm inv}(\pi^0 \Lambda(1405)), m_{\pi^0}^2, M^2_{\rm inv}(\pi^0 \Sigma^0))}{2 M_{\rm inv}(\pi^0 \Lambda(1405))}.
\end{equation}
We follow the method of Ref. \cite{Pavao:2017kcr}, and obtain the final differential  distribution for four particles in the final state,
\begin{widetext}
\begin{equation}\label{eq:dG1}
\frac{1}{\Gamma_{\Lambda_c^+}}\frac{d^2 \Gamma}{d M_{\text{inv}}(\pi^0  \Lambda(1405)) d M_{\text{inv}}(\pi^0 \Sigma^0)}  = \frac{1}{(2 \pi)^5} \frac{M_{\Sigma^0}}{M_{\Lambda_c^+}}
\widetilde{p}_{\pi^+} \widetilde{p}_{\pi^0} \widetilde{q}_{\Sigma^0} \frac{1}{2} g^2 \frac{A^2}{\Gamma_{\Lambda_c^+}} |\vec{k}|^2 \left| t_T\right|^2  \times \left| t_{K^- p \rightarrow \pi^0 \Sigma^0}\right|^2,
\end{equation}
\end{widetext}
with
\begin{subequations}
\begin{align}
&\widetilde{p}_{\pi^+} = \frac{\lambda^{1/2} (M_{\Lambda_c^+}^2, M^2_{\text{inv}}(\pi^0 \Lambda(1405)),m^2_{\pi^+})}{2 M_{\Lambda_c^+} },	\\
&\widetilde{p}_{\pi^0}=|\vec{k}|= \frac{\lambda^{1/2} ( M^2_{\text{inv}}(\pi^0 \Lambda(1405)),m^2_{\pi^0}, M^2_{\text{inv}}(\pi^0 \Sigma^0)}{2 M_{\text{inv}}(\pi^0 \Lambda(1405))},\\
&\widetilde{q}_{\Sigma^0}=\frac{\lambda^{1/2} ( M^2_{\text{inv}}(\pi^0 \Sigma^0),m^2_{\pi^0}, M^2_{\Sigma^0})}{2 M_{\text{inv}}(\pi^0 \Sigma^0)}.
\end{align}
\end{subequations}

To regularize the integral in Eq.~\eqref{eq:tt2} we use the same cutoff of the meson loop that is used to calculate $t_{K^- p \to \pi^0\Sigma^0}$ in Eq.~\eqref{eq:vertex3} with $\theta(q_{\text{max}}-|\vec{q}^*|)$, where $\vec{q}^{\ *}$ is the $\vec{q}$ momentum in the $R$ rest frame (see Ref.~\cite{Bayar:2016ftu}).

\subsection{Isospin-breaking effect}

If we use the same masses for $\bar{K}N$ and $K^*$ with isospin conservation, we find that the contributions from  Fig. \ref{fig:LcpiR}~(a) and Fig. \ref{fig:LcpiR}~(b) will cancel each other. An interesting thing is to investigate the isospin-breaking effect.  That means, for the first time,  we will precisely look at the $\Lambda(1405)$ formation in an isospin forbidden mode.  We expect that the formation will be driven by a triangle singularity and the shape will be narrower than usual, because it will be tied to the different masses of $\bar{K} N$.  Therefore,  in the following, we will use different masses for  $K^- p$ or $ \bar{K}^0 n$, and also for  $K^{*-}$ and $\bar{K}^{*0}$.

Now we consider Fig. \ref{fig:LcpiR}~(b),  for the  $\Lambda_c^+ \to \pi^+ \bar{K}^{*0} n $ followed by  $\bar{K}^{*0} \to \pi^0 \bar{K}^0$ decay and
$\bar{K}^0 n \to \pi^0\Sigma^0$
to see the $\Lambda(1405)$ formation.  We  also need to calculate the three vertices, $t_{\Lambda_c^+ \to \pi^+ \bar{K}^{*0} n} $, $t_{\bar{K}^{*0} \to \pi^0 \bar{K}^0}$ and $t_{\bar{K}^0 n \to \pi^0\Sigma^0}$.

{For the first vertex,
$t_{\Lambda_c^+\rightarrow\pi^+\bar{K}^{*0}n}$, we can use the same
amplitude in Eq.~\eqref{eq:A}
\begin{align}
 t_{\Lambda_c^+\rightarrow\pi^+\bar{K}^{*0}n}=A\vec{\sigma}\cdot\vec{\epsilon}.
\end{align}
As shown in Eq.~\eqref{h}, the weight of the production is the same as
the $\pi^+K^{*-}p$ process with the same sign.
Then, we can use the same $A$ in this case.

The amplitude $t_{\bar{K}^0\rightarrow\pi^0\bar{K}^0}$ for the second vertex
is written by using Eq.~\eqref{eq:vpp} as two times as $K^{*-}\rightarrow\pi^0K^-$
\begin{align}
 -it_{K^{*0}\rightarrow\pi^0\bar{K}^0}=&-ig\frac{1}{\sqrt{2}}\epsilon^\mu_{\bar{K}^{*0}}(p_{\bar{K}^0}-p_{\pi^0})\\
 \simeq&-ig\frac{1}{\sqrt{2}}\vec{\epsilon}_{\bar{K}^{*0}}\cdot(\vec{p}_{\pi^0}-\vec{p}_{\bar{K}^0}),
\end{align}
where the amplitude has the opposite sign to the
$\bar{K}^{*-}\rightarrow\pi^0K^-$ in Eq.~\eqref{eq:v21a}.
Finally, the $\bar{K}^0n\rightarrow\pi^0\Sigma^0$ amplitude for the
third vertex $t_3=t_{if}$ is the component with $i=2$ and $f=4$.

The triangle amplitude for the $\bar{K}^{*0}n\bar{K}^0$ loop,
$t_T=t_T(m_{\bar{K}^{*0}},M_n,m_{\bar{K}^0})$, is obtained with
Eq.~\eqref{eq:tt2} replacing the masses and width of the internal particles.}

 Hence, for the isospin-breaking effect,  we get the final differential  distributions,
\begin{widetext}
\begin{align}\label{eq:dG2}
\frac{1}{\Gamma_{\Lambda_c^+}}\frac{d^2 \Gamma}{d M_{\text{inv}}(\pi^0  \Lambda(1405)) d M_{\text{inv}}(\pi^0 \Sigma^0)}= \frac{1}{(2 \pi)^5} \frac{M_{\Sigma^0}}{M_{\Lambda_c^+}} \widetilde{p}_{\pi^+} \widetilde{q}_{\Sigma^0} \frac{1}{2} g^2
\frac{A^2}{\Gamma_{\Lambda_c^+}} |\vec{k}|^3 \\\nonumber
\times \left| t_T(m_{K^{*-}},M_p,m_{K^-}) t_{K^- p \to \pi^0 \Sigma^0}- t_T(m_{\bar{K}^{*0}},M_n,m_{\bar{K}^0}) t_{\bar{K}^0 n \to \pi^0 \Sigma^0} \right|^2,
\end{align}
\end{widetext}

\section{Results}

\begin{figure}
	\begin{center}
		\includegraphics[width=0.68\textwidth]{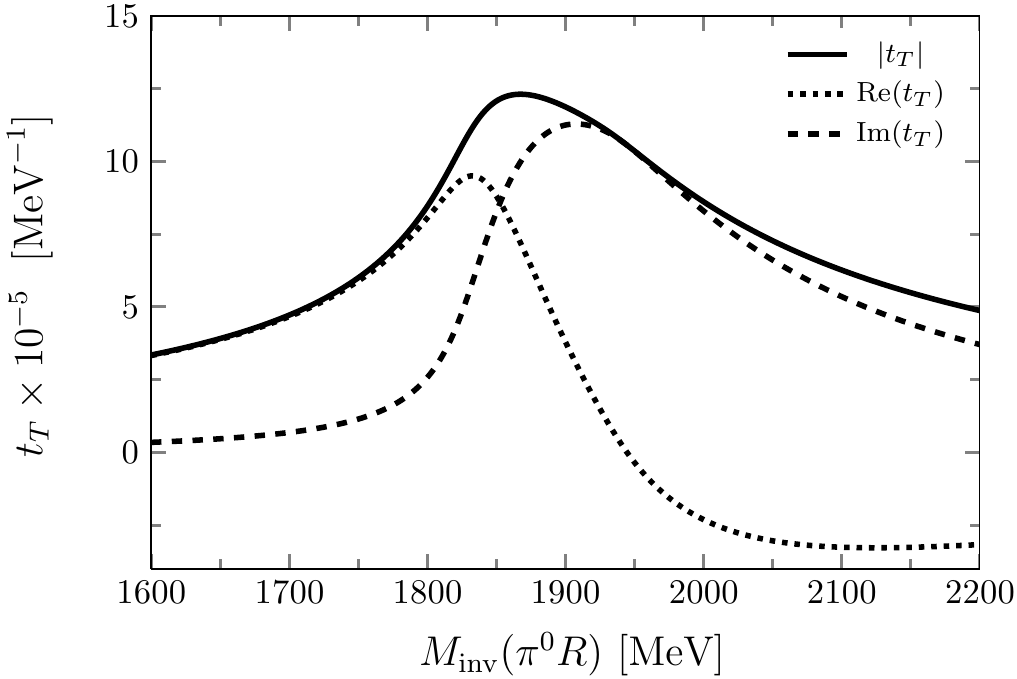}
	\end{center}
	\caption{\label{Fig:tT} Triangle amplitude $t_T$ for the decay in  Fig. \ref{fig:LcpiR}~(a), here taking $M_{\rm inv}(R)$=1420 MeV.}
\end{figure}

Let us begin by showing in Fig. \ref{Fig:tT} the contribution of the triangle loop defined in Eq. \eqref{eq:tt2}.
We plot the real and imaginary  parts of $t_T(m_{K^{*-}},M_p,m_{K^-})$, as well as the absolute value  with $M_{\rm inv}(R)\equiv M_{\rm inv}(\pi^0 \Sigma^0) $ fixed at 1420 MeV.
It can be observed that the ${\rm Re}(t_T)$  has a peak around 1838 MeV, and  ${\rm Im}(t_T)$  has a peak around 1908 MeV,
and  there is a peak for $|t_T|$ around 1868 MeV. As discussed in Ref.~\cite{Sakai:2017hpg}, the peak of the real part is related to the $K^{*-} p$ threshold and the one of
the imaginary part,  that dominates for the larger $\pi^0 R$ invariant masses, to the triangle singularity.

\begin{figure}
	\begin{center}
		\includegraphics[width=0.68\textwidth]{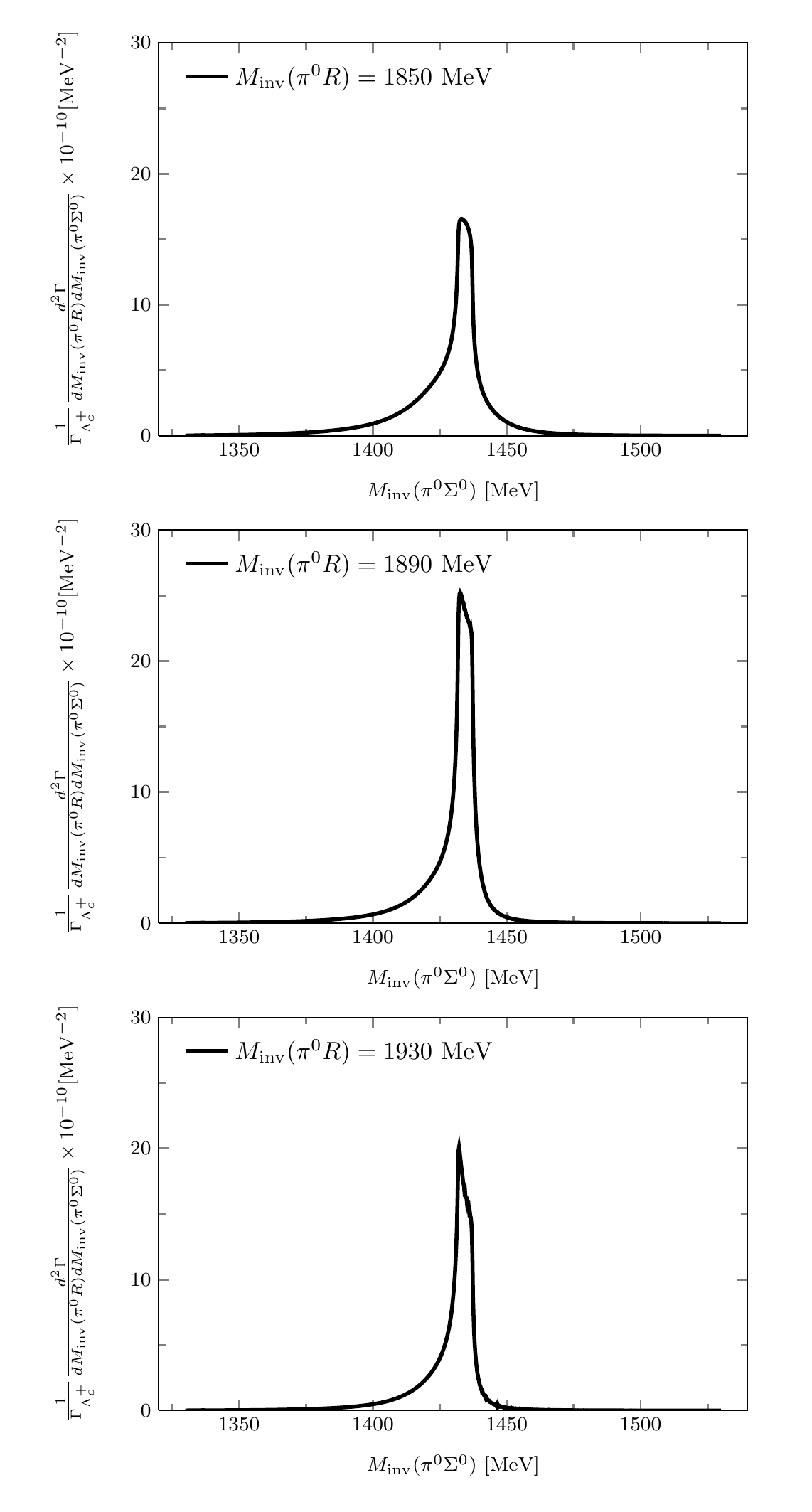}
	\end{center}\vskip -1.cm
	\caption{\label{Fig:dGdMM}The derivative of the mass distribution of $\Lambda_c^+ \to \pi^+\pi^0\pi^0\Sigma^0$  with regards to $M_{\rm inv}(\Lambda(1405))$.}
\end{figure}

In Fig.~\ref{Fig:dGdMM} we plot Eq. \eqref{eq:dG2} for $\Lambda_c^+ \to \pi^+\pi^0\pi^0\Sigma^0$ by
fixing $M_{\rm inv}(\pi^0 R)$=1850 MeV, 1890 MeV, and 1930 MeV  and varying  $M_{\rm inv}(R)$.  We can see that the distribution with largest strength is near $M_{\rm inv}(\pi^0 R)$=1890 MeV.
We can also see a strong peak around  1432 MeV for the three different masses of $M_{\rm inv}(\pi^0 R)$. Consequently, we see that most of the contribution to our width $\Gamma$ will come from $M_{\rm inv}(R)=M_R$,
thus we have strong contributions for $M_{\rm inv}(\pi^0 \Sigma^0 ) \in [1390~\rm{MeV},1450~\rm{MeV}]$.  The conclusion is that when we
calculate the mass distribution $\frac{d \Gamma}{d M_{\rm inv}(\pi^0 \Lambda(1405))}$, we can restrict the integral in  $M_{\rm inv}(R)$ to the limits already mentioned.

\begin{figure}
	\begin{center}
		\includegraphics[width=0.68\textwidth]{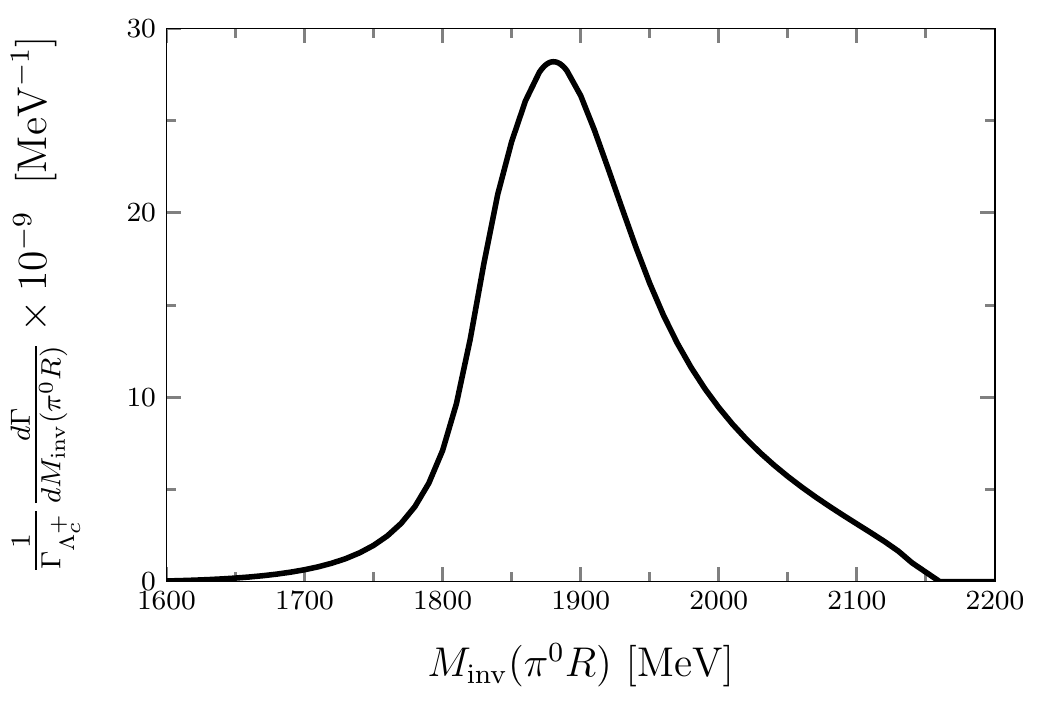}
	\end{center}
  \caption{\label{Fig:dgdm} The mass distribution of  $\Lambda_c^+ \to \pi^+\pi^0\pi^0\Sigma^0$.}
\end{figure}

By integrating over $M_{\rm inv}(R)$, we obtain $\frac{1}{\Gamma_{\Lambda_c^+}}\frac{d \Gamma}{d M_{\rm inv}(\pi^0 R)}$ which is  shown in Fig.~\ref{Fig:dgdm}.
We see a clear peak of the distribution around 1880 MeV for $\Lambda(1405)$ production.
When performing the integral we observe that the strong contribution comes from $M_{\rm inv}(\pi^0 \Sigma^0 ) \in [1390~\rm{MeV},1450~\rm{MeV}]$.  The conclusion is that when we
calculate the mass distribution $\frac{d \Gamma}{d M_{\rm inv}(\pi^0 \Lambda(1405))}$, we can restrict the integral in  $M_{\rm inv}(\pi^0\Sigma^0)$ to the limits already mentioned.

Integrating now $\frac{d \Gamma}{d M_{\rm inv}(\pi^0 \Lambda(1405))}$ over  $M_{\rm inv}(\pi^0 \Lambda(1405)) \in [1800~\rm{MeV}, 2050~\rm{MeV}]$   in Fig.~\ref{Fig:dgdm}, we obtain the branching fraction
\begin{equation}
 Br(\Lambda_c^+ \rightarrow \pi^+ \pi^0 \Lambda(1405); ~~\Lambda(1405) \rightarrow \pi^0 \Sigma^0) = ( 4.17 \pm  1.39) \times 10^{-6}.
\end{equation}
This  number is within measurable range.  The errors come from the experimental errors in the branching ratio of $Br(\Lambda_c^+ \to  \pi^+ K^{*-}  p)$.

One should stress the most remarkable feature in the distributions of  Fig.~\ref{Fig:dGdMM}: the width of the $\Lambda(1405)$ produced is a mere 6.5 MeV, remarkably smaller than the nominal widths for the
$\Lambda(1405)$  at 1420 MeV of the order  of 30 MeV. As mentioned before, this narrow width is tied basically to the different masses of the $K^-,\bar{K}^0$ or $p,n$.  This exceptionally narrow shape has been
observed in all the isospin  forbidden $f_0(980)$ production mode. The present reaction  would be the first one where  the narrow  $\Lambda(1405)$ is seen in an isospin forbidden mode.

\section{Conclusions}
\label{sec:conc}
 The triangle singularities have recently shown to be very effective, enhancing the production of  $f_0(980)$ or $a_0(980)$ in isospin suppressed modes. These reactions have played a double role. On the first hand they have provided clear examples of triangle singularities, and on the second hand the rates and shapes obtained for the isospin suppressed modes are closely tied to the nature of these resonances and offer extra support to their dynamical origin from the interaction of mesons in coupled channels.

    The two states of the $\Lambda(1405)$, now already official in the PDG, are another example of dynamical generation from the interaction of meson-baryon in this case. Yet, the resonance has not been observed in an isospin violating reaction so far. The present work provides the first evaluation of the $\Lambda(1405)$ production in an isospin forbidden reaction. We devised one such reaction, which, as in the case of the $f_0(980)$ or $a_0(980)$, can be enhanced by a triangle singularity. We found such an example in the decay of $\Lambda_c^+$ into $\pi^+ \pi^0 \Lambda(1405)$. The mechanism for the production is given by a first decay of the $\Lambda_c^+$ into $\pi^+ \bar K^*  N $, then the $\bar{K}^*$ decays into $\bar K \pi $ and the $\bar K N$ merge to
 produce the  $\Lambda(1405)$ through a triangle loop containing $\bar{K}^* N \bar K$, which develops a singularity around 1890 MeV.

    The remarkable observation is that a peak tied to the $\Lambda(1405)$ state of higher energy (around 1420 MeV) appears in the final $\pi^0 \Sigma^0$ mass spectrum, but peaking even at higher energy, close to the $\bar K N$ threshold of 1432 MeV. It is also remarkably narrow, of the order of 6-7 MeV, and is tied to the difference of masses between the $K^-$ and $\bar K^0$ and $p, n$.

   We have shown that the amount of  $\Lambda(1405)$ production has its largest strength at a $\pi^0 \Lambda(1405)$ invariant mass of around 1890 MeV, where the mechanism suggested develops a triangle singularity. The shape and strength obtained are intimately tied to the nature of the $\Lambda(1405)$ as a dynamically generated resonance from the meson baryon interaction, and in the present case, to its
large coupling  to the $\bar K N$ component.  We found that the strength of the width observed falls within measurable range. The implementation of the reaction would thus bring valuable information on the nature of this resonance, the mechanisms of triangle singularities, plus extra information on the continuously searched for $\bar K N$ interaction.

\section*{Acknowledgements}
L. R. Dai wishes to acknowledge the support from the National Natural Science Foundation of China (NSFC)
(No. 11575076) and the State Scholarship Fund of China (No. 201708210057).  R. P. Pavao wishes to
thank the Generalitat Valenciana in the program Santiago Grisolia.
This work is partly supported by the Spanish Ministerio de Economia y Competitividad
and European FEDER funds under the contract number FIS2011-
28853-C02-01, FIS2011- 28853-C02-02, FIS2014-57026-
REDT, FIS2014-51948-C2- 1-P, and FIS2014-51948-C2-
2-P, and the Generalitat Valenciana in the program
Prometeo II-2014/068 (EO).

\end{document}